\RequirePackage{lineno}
\documentclass[aip, amsmath,amssymb, reprint]{revtex4-1}
\usepackage{graphicx}
\usepackage{dcolumn}
\usepackage{bm}
\usepackage{verbatim}
\usepackage{xcolor}
\usepackage{braket}
\usepackage{pstricks-add}
\usepackage{tikz}
\usetikzlibrary{graphs}
\usepackage{subfigure}
\usepackage[utf8]{inputenc}
\usepackage[T1]{fontenc}
\usepackage{mathptmx}
\usepackage{etoolbox}
\usetikzlibrary{decorations.pathmorphing} 
\usetikzlibrary{matrix} 
\usetikzlibrary{arrows} 
\usetikzlibrary{calc} 
\tikzstyle{snakeline} = [decorate, decoration={pre length=0.1cm,
                         post length=0.1cm, snake, amplitude=0.3mm,
                         segment length=1.5mm},thick, magenta, ->]

\usepackage{CJK}

\makeatletter
\def\@email#1#2{%
 \endgroup
 \patchcmd{\titleblock@produce}
  {\frontmatter@RRAPformat}
  {\frontmatter@RRAPformat{\produce@RRAP{*#1\href{mailto:#2}{#2}}}\frontmatter@RRAPformat}
  {}{}
}%
\makeatother

\begin{document}


\begin{CJK*}{UTF8}{gbsn}
\title{New Opportunities in  Femto-to-Nanometer Scale with High-Intensity Lasers }

\def\FDU{Key Laboratory of Nuclear Physics and Ion-beam Application (MoE), Institute of Modern Physics, Fudan University, Shanghai 200433,  China}
\def\sjtu {Department of Physics and Astronomy, Shanghai Jiao Tong University, Shanghai, 200240, China}
\def\iop  {Lab of Optical Physics, Institute of Physics, Chinese Academy of Sciences, Beijing 100190, China}
\def\imp {Institute of Modern Physics, Chinese Academy of Sciences, Lanzhou, 730000, China}
\def\ciae {Department of Nuclear Physics, China Institute of Atomic Energy, Beijing, 102413, China}
\def\nao {National Astronomical Observatories, Chinese Academy of Sciences, Beijing 100012, China}
\def\ifsa {IFSA Collaborative Innovation Center, Shanghai Jiao Tong University, Shanghai 200240, China}
\def\siom{Shanghai Institute of Optics and Fine Mechanics, Chinese Academy of Sciences, Shanghai 201800, China}
\def\CIAE{China Institute of Atomic Energy, Beijing 102413, China}
\def\ihep{Institute of High Energy Physics, CAS, Beijing, 100049, China}
\def\bnu{College of Nuclear Science and Technology, Beijing Normal University, Beijing, 100012, China}
\def\SARI{Shanghai  Advanced Research Institute, Chinese Academy of Sciences, Shanghai 201210, China}
\def\IAP{Institute of Applied Physics and Computational Mathematics, Beijing, 100094, China}

\author{Changbo Fu}\email[Email:] {cbfu@fudan.edu.cn} \affiliation{\FDU}
\author{Guoqiang Zhang} \email[Email:] {zhangguoqiang@zjlab.org.cn} \affiliation{\SARI}
\author{Yugang Ma} \email[Email:] {mayugang@fudan.edu.cn}\affiliation{\FDU}


\date{\today}

\begin{abstract}
On the scale of nanometer to femtometer, there have several puzzles, 
including neutron lifetime, proton charge radius, and deep Dirac level, etc.
With the development of high-intensity laser technologies, 
lasers today can induce extremely strong electromagnetic (EM) fields.
Electrons in deep shells of atoms, as well as the atomic nucleus themself, can be affected by the laser EM fields.
This may provide a new experimental platform for studies the physical processes on the femto-to-nanometer scale,
where atomic physics and nuclear physics coexist.
In this paper, we review possible new opportunities for studying the puzzles on femto-to-nanometer scale with high-intensity lasers.
\end{abstract}
\maketitle

\maketitle

\section{Introduction}
%
Atomic physics, which has a typical scale of nanometers, and nuclear physics, 
which has a typical scale of femtometers, have been developed over 100 years. 
The $\alpha$ particle backscattering experiment  performed by Ernest Rutherford in 1909 led to the birth of the Nuclear Model of an atom\cite{QM-Schiff}. 
It is a major step towards how we see the atom today:
a nucleus with a diameter of femtometer order covers by electron clouds which has a diameter of nanometer order.
It may look like that after over 100 years physical phenomena on femto-to-nanometer scale (FNMS) should have been understood well.
However, the facts are far from that. 

On FNMS, the laws of both atomic physics and nuclear physics play important roles.
On FNMS, forces involved are relatively simple compared with those in nuclei. 
Beyond a nuclear surface, the strong force decreases very quickly,
and normally only the electromagnetic force dominates the dynamics of electrons. 
However, in some cases, if a nucleus is radioactive, the weak force can also play an important role.
In spite of the simple forces, on FNMS, there still have many phenomena which make the physics communities confused. 
The puzzles include the proton charge radius puzzle\cite{P-size-Karr-Nat-Rev2020}, the neutron decay lifetime puzzle\cite{n-lifetime-PRL2013Yue,n-time-RMP2011}, and the Deep Dirac Level puzzle\cite{Fu.DDL.NPR.2020}, etc. 
Studies of these puzzles may result in deeper understanding of material structures, or even new physics beyond the Standard Model.

The development of high-intensity laser technologies provides a unique way for studies on physics on FNMS.
Without any doubt, lasers are very useful tools for studying microcosmos.   
One can deduce structures of molecules and atoms by using narrowband lasers\cite{Surf-Raman-PRL1997Kneipp}.
In temporal scales, one can deduce  chemical bond dynamics by using ps and/or fs lasers\cite{Laser-Chem-2019}.
Because of the electron clouds, nuclei are well shielded and are not so intensively considered.
With high-intensity lasers, atoms can be highly ionized, and importance of nuclei shows up. 
However, the new opportunities on FNMS, which connect atomic physics and nuclear physics, 
have not been fully discussed yet.

In this paper, we will review several puzzles which exist on FNMS first. 
And then review several theoretical tools on FNMS which can be used to analyze many-body systems involving photons (laser), electrons, and nuclei. 
The applications like nuclear clocks and nuclear batteries will be also discussed.

\section{Puzzles on FNMS}

\subsection{Neutron lifetime puzzle}

A free neutron will decay into a proton, electron, and antineutrino through the weak interaction,
$n\rightarrow p + e + \bar{\nu}_e$.
The neutron decay lifetime is very important to fields like particle physics and astrophysics\cite{bbn-2020Fields}.
In particle physics, the neutron lifetime plays a critical role in determining basic parameters like quark mixing angles,   
quark coupling constant, and cross sections related to weak p-n interactions\cite{n-puzz-2009PAUL,NeutrinoCooling2019CPL}.
In the nuclear astrophysics field, neutron lifetime determines the speed of nucleosynthesis in the Big Bang and in stars.
In the few seconds after the Big Bang, protons and neutrons formed.
And a few minutes later, as the universe expands, 
the temperature of the universe dropped below the photo-dissociation threshold for deuterons,
the equilibrium ratio of proton and neutron broken,
and then the primordial nucleosynthesis starts.
A precise neutron lifetime is an important input parameter for the primordial nucleosynthesis calculations.

It is very interesting that neutron lifetime has not had a consistent experimental value yet.
Different neutron lifetime values were obtained by different groups which used different techniques. 

There have two major types of experiments to measure the neutron lifetime, 
beam experiments which use neutron fluxes,
and bottle experiments which use ultra-cold neutrons confined in containers\cite{n-time-RMP2011,n-lifetime-PRL2013Yue}.
After many years of efforts, 
the beam method measures the lifetime  as $(888.1\pm2.0)$ s,
while the bottle method $(879.45\pm0.58)$ s\cite{n-lifetime-atoms2018}.
The difference between the averages from the two methods, beam and bottle experiments,   is $(8.7 \pm 2.1)$ s,
which is  $4.1\sigma$.  
This persistent disagreement may relate to an unknown process in neutron decay,
or even relate to science beyond the Standard Model.

The neutron decay $n\rightarrow p + e + \bar{\nu}_e$ involves the weak interaction,
and the weak interaction has effects both in the nuclear scale and atomic scale, 
or in other words, the FNMS. 
A deeper study on FNMS would help to resolve the neutron lifetime puzzle.


\subsection{Deep Dirac Level puzzle}

The so-called deep Dirac level (DDL) is another very interesting puzzle existing on FNMS. 
The history of the DDL can be traced back to the time when the Dirac equation was established for the first time.
It is well known today that some ``unphysical solution" was explained as positron solutions.
In fact, there still has another ``unphysical solution'', the DDL, which is not so well known, was normally rejected, 
but keeping attracting attention from different physical fields. 
Many people suspect that this ``unphysical solution" DDL may relate to a physical state just like that in the ``positron solution'' case.
 
For an electron moving around a nucleus, the corresponding Dirac equation can be written as\cite{DiracEq-1997-Goodman},
\begin{equation}\label{eq.Dirac}
(\bm{\alpha\cdot p}+\beta M +V(r)) \bm{\Psi}=E\bm{\Psi}
\end{equation}
where the Dirac matrices are
\begin{equation}\label{eq.a.b}
\bm{\alpha}= 
\begin{pmatrix}
0 & \bm{\sigma} \\
\bm{\sigma}  &0
\end{pmatrix},
\beta= 
\begin{pmatrix}
I & 0 \\
0  & -I 
\end{pmatrix},
\end{equation}
$\bm{\sigma}$ is Pauli matrix, $I$ is unit matrix,
$V(r)=-\frac{Z \alpha\hbar c}{r}$ is the coulomb potential, 
$\alpha$ is the fine structure constant, 
$r$ is distance between electron and nuclei, 
and $Z$ is the charge of the nuclear.
The Dirac equation Eq.\ref{eq.Dirac} has solution of, 
\begin{equation}
\bm{\Psi}= 
\begin{pmatrix}
g(r)\Omega_{jlm}(\theta,\phi) \\
i f(r)\Omega_{j\bar{l}m}(\theta,\phi) 
\end{pmatrix},
\ j=l\pm \frac{1}{2},\ l+\bar{l}=2j,
\end{equation}
$\Omega_{jlm}$ are two-component angular momentum spinors.
With ansatz of $f\propto r^\gamma$ and $g\propto r^\gamma$, one has \cite{QM-Schiff}
\begin{equation}
\gamma=\pm\sqrt{\left(j+1/2\right)^2-Z^2\alpha^2}.
\end{equation}
It is argued that the expectation value of the Coulomb energy,
\begin{eqnarray}\label{eq.coulomb.energy.int}
E_C&=&\int \Psi^\dagger \left( -\frac{Z e^2}{r} \right)\Psi {\rm d}\vec{r}\nonumber\\
&=&\int (f^2+g^2) \left( -\frac{Z e^2}{r} \right) {\rm d}\vec{r}\nonumber\\
&\propto&\int r^{2\gamma} \left( -\frac{Z e^2}{r} \right) {\rm d}\vec{r}.
\end{eqnarray}
For a negative $\gamma$, one has  $ E_C|_{\gamma<-1/2}\rightarrow\infty$, which is unphysical.
However, this infinity comes from the point-like potential $V(r)=-\frac{Z \alpha\hbar c}{r}$. i.e. the assumption that the charge of the nucleus is a point. 
If considering the nucleus has a limited charge radius, 
the singularity is removed and the $E_C$ is limited too.
The question turns to be that the solutions for nucleus inside and outside should be connected on the nucleus surface  \cite{QM-Greiner}.

One can even see the unordinary solution more clearly from the Klein-Golden equation.
It is well known that the Dirac equation can be transferred to  KG equation by taking the square of both sides of Eq.\ref{eq.Dirac},
\begin{equation}
\left[
\bigtriangledown^2+(V(r)-i\hbar\partial_t)^2-M^2
\right]
\bm{\Psi}=0.
\end{equation}

In spherical coordinates, it can be written as,
\begin{eqnarray}
&&\left[
\frac{1}{r}\frac{\partial}{\partial r}\left(r^2 \frac{\partial}{\partial r}\right)
+\frac{1}{r^2\sin \theta}\frac{\partial}{\partial \theta}\left(\sin\theta \frac{\partial}{\partial \theta}\right) \right. \nonumber\\
&&\left. +\frac{1}{r^2\sin^2 \theta}\frac{\partial^2}{\partial \phi^2}
+(i\hbar\partial_t-V)^2-M^2
\right]
\psi(r,\theta,\phi)=0. \nonumber \\
\end{eqnarray}
The total wave function $\psi(r,\theta,\phi,t)$ can be assigned as,
\begin{equation}\label{eq.psi.r.t.p.t}
\psi(r,\theta,\phi,t)=\frac{R(r,t)}{r}Y_{lm}(\theta,\phi)=\frac{R(r,t)}{r}\Theta(\theta)\Phi(\phi).
\end{equation}
$R(r,t)$, $\Theta(\theta)$, and $\Phi(\phi)$ are solutions of the equations
\begin{eqnarray}\label{eq.Dirac.R}
\left[\frac{{\rm d}^2}{{\rm d} r^2}+(i\hbar\partial_t-V)^2-M^2 -\frac{l(l+1)}{r^2}\right]R(r,t)&=&0, \nonumber\\
\frac{{\rm d}^2 \Theta(\theta)}{{\rm d} \theta^2}
+\cot\theta\frac{{\rm d}\Theta(\theta)}{{\rm d}\theta}\left[l(l+1) -\frac{m^2}{\sin^2\theta}\right]\Theta(\theta)&=&0, \nonumber\\
\frac{{\rm d}^2 \Phi(\phi)}{{\rm d} \phi^2}
+m^2\Phi(\phi)&=&0, \nonumber \\
\end{eqnarray}
respectively.

In the case of $l=0$, the Eq.\ref{eq.Dirac.R} has solution of\cite{naudts2005hydrino-KG-eq}, 
\begin{equation}\label{eq.psi.t}
R(r,t)=\frac{r_0^{-\gamma/2-1}  r^{(\gamma-1)/2} e^{-r/r_0}}{2^{-\gamma/2}\sqrt{\pi \Gamma(\gamma+2)}}e^{-iE_0 t/\hbar},
\end{equation}
where $\Gamma(x)$ is the gamma function,
$E_0=m_0 c^2\frac{Z\alpha}{\sqrt{(1-\gamma)/2}}$,
and $r_0=\frac{\hbar}{m_0c}\frac{1}{\sqrt{(1-\gamma)/2}}$.

The same as the Dirac equation case, the negative branch  $\gamma=-\sqrt{1-4Z^2\alpha^2}$  solution is the ``unphysical'' one\cite{QM-Schiff}. 
The energy, orbit, and wave function can be simplified as,
\begin{equation}
E_0^\#=m_0 c^2\cdot Z\alpha\simeq m_0 c^2-(511-3.72\cdot Z) keV,
\end{equation}
\begin{equation}
r_0^\#\simeq\frac{\hbar}{m_0c}\simeq 0.0039\, \overset{\circ}{A}.
\end{equation}
\begin{equation}\label{eq.psi.ddl}
R^\#(r,t)\simeq\frac{ r^{-1}e^{-r/r_0^{\#}}}{\sqrt{2\pi r_0^{\#}}} e^{-i E_0^\# t/\hbar},
\end{equation}

It can be found that $\lim\limits_{r \to 0}R^{\#} =\infty$. 
However, the singularity can be removed if the nucleus is not a point-like particle as assumed in  $V(r)=-Z \alpha\hbar c/r$. 
The wave function itself is also square integrable, i.e. $\int |\Psi|^2 {\rm d}\vec{r}=1$.
Furthermore, one can check that the $E_C$ in Eq.\ref{eq.coulomb.energy.int} is also finite if insert Eq.\ref{eq.psi.r.t.p.t} and \ref{eq.psi.ddl} into it. 

It is worth noticing that a solution of a Klein-Golden equation is not automatically turning to be a solution of the corresponding Dirac equation.
An experimental discovery of an electron's state corresponding to the Klein-Golden equation, but not to the Dirac equation, 
may give hints that the electron is under symmetry broken between fermion and boson.

The existence of the DDLs is debated theoretically, as well as experimentally. Some experimental phenomenon have been explained as existence of the DDLs\cite{phillips2004water2004,mills2003extreme, va2013new}, 
but being questioned by others\cite{jovicevic2009spectroscopic, rathke2005critical, phelps2005comment, dombey2006hydrino, de2007orthogonality}.  
New sensitive experimental detecting method is needed. 
 
As one can see, taking $Z = 1$ as an example, in the case of the DDL, the electron is deeply bounded to 0.5073 MeV, compared with the well-known Bohr case 13.6 eV. 
The DDL’s orbit is only about 390 fm away from the nucleus, compared with the Bohr orbit, 
$0.53 \overset{\circ}{A}$. 
In \cite{Fu.DDL.NPR.2020}, the authors proposed the electron capture lifetime as a novel indicator of DDL existing, for the first time. 
Because the DDL orbit is much closer to the nuclei, the probability of being captured is dramatically enhanced.

\subsection{Proton charge diameter puzzle}


Neutrons and protons are fundamental building blocks of the visible matter around us.
However, even after being discovered over 100 years (discovery of the neutron credited to  Jame Chadwick in 1932, and proton  Ernest Rutherford in 1917), 
there are still some basic physical quantities we do not know yet. 

 The proton charge radius has been measured since the 1950s\cite{P-size-Karr-Nat-Rev2020,size-p-Nature-2020Pohl}.
 After over  scientific research, the proton is still mysterious in many ways. 
 
 The proton charge radius can be determined using two major experimental techniques, 
 the electron–proton elastic scattering,
 and the high-resolution spectroscopy of electronic and muonic hydrogen atoms.
The different techniques result in conflict proton charge radius.
Over ten different electronic transitions have been measured in electronic hydrogen, as well as in muonic hydrogen\cite{P-size-Karr-Nat-Rev2020}. 
Most electronic hydrogen results are compatible with the muonic hydrogen ones within 1.5$\sigma$.
For the electron-proton elastic scattering, 
the results from different groups differ as large as 5$\sigma$\cite{P-size-Karr-Nat-Rev2020}.
The conflict is a prime problem in proton structure physics.
The solving of this conflict may result in a new discovery in proton and hydrogen atom structure on FNMS.

To resolve the disagreement between the results,
one may need a better understanding of the structure on FNMS.
For example, if DDL exists, the two processes, 
the e-p scattering and the lamb-shift of the hydrogen atoms could be different.
On the one hand, a precise proton radius value would be input to solve other puzzles on FNMS;
on the other hand, new structures like DDL may help to resolve the proton size puzzle.

\section{Mechanisms on FNMS}

\subsection{NEEC and NEET}
\begin{figure}[t]
 \centering

\begin{tikzpicture}[scale=0.2]
 \pgfmathsetmacro{\L}{8}
 \pgfmathsetmacro{\H}{6}
 \pgfmathsetmacro{\Gap}{1.5}
 \draw[thick, black] (0,0) -- (\L,0);
 \draw[thick, blue] (\L+\Gap,0) -- (2*\L+\Gap,0);
 \draw[thick, black] (2*\L+5*\Gap,0) -- (3*\L+5*\Gap,0);
 \draw[thick, blue,dotted] (3*\L+6*\Gap,0) -- (4*\L+6*\Gap,0);
 \draw[thick, fill=white,font=\fontsize{6}{6}\selectfont] (2.5*\L+5*\Gap,0) circle (0.5) node {e};
 \draw[thick, fill=red,font=\fontsize{6}{6}\selectfont] (1.5*\L+1*\Gap,0) circle (0.6) node {N};

 \draw[very thick, blue,dotted] (2*\L+3*\Gap,-0.5*\H) -- (2*\L+3*\Gap,3*\H);
	
 \draw[thick, blue,dashed,font=\fontsize{6}{6}\selectfont] (-0.2*\L,1.3*\H) --  node [above]{unbounded} (\L+0.8*\Gap,1.3*\H);
 \draw[thick, blue,dashed,font=\fontsize{6}{6}\selectfont] (-0.2*\L,1.3*\H) --  node [below]{bounded}     (\L+0.8*\Gap,1.3*\H);

 \draw[thick, blue,dashed,font=\fontsize{6}{6}\selectfont] (2*\L+4*\Gap,1.3*\H) --  node [above]{unbounded} (3*\L+6*\Gap,1.3*\H);
 \draw[thick, blue,dashed,font=\fontsize{6}{6}\selectfont] (2*\L+4*\Gap,1.3*\H) --  node [below]{bounded}     (3*\L+6*\Gap,1.3*\H);
 
 \draw[thick, blue] (\L+\Gap,2*\H) -- (2*\L+\Gap,2*\H);
 \draw[thick, blue] (3*\L+6*\Gap,2*\H) -- (4*\L+6*\Gap,2*\H);
 \draw[thick, fill=white, font=\fontsize{6}{6}\selectfont] (\L/2,2*\H) circle (0.5) node {e} node[left]{free};
 \draw[thick, fill=red,font=\fontsize{6}{6}\selectfont] (3.5*\L+6*\Gap,2*\H) circle (0.6) node {N};
 \draw[snakeline] (\L/2, 2*\H-0.6) -- (0.5*\L,0);
 \draw[snakeline] (1.5*\L+\Gap, 0.6) -- (1.5*\L+\Gap,2*\H);

 \node at (0.5*\L,2.6*\H) {Atomic};
 \node at (2.5*\L+5*\Gap,2.6*\H) {Atomic};
 \node at (1.5*\L+\Gap,2.6*\H) {Nuclear};
 \node at (3.5*\L+6*\Gap,2.6*\H) {Nuclear};

 \node at (\L+0.5*\Gap,-0.6*\H) {Initial State};
 \node at (3*\L+6.5*\Gap,-0.6*\H) {Final State};

 \end{tikzpicture}	

\caption{
Atomic and nuclear states involved in a NEEC transition.
In the initial state, the electron is unbounded and the nucleus on its ground state;
while in the final state, the electron is bounded and the nucleus is excited to a high energy state.
}

 \label{fig.NEEC}
\end{figure}
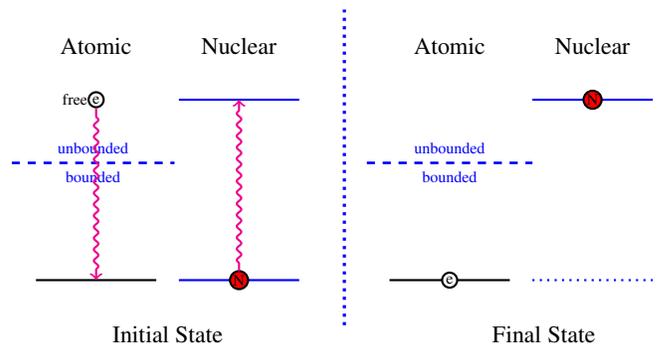

Internal conversion is a well-known nuclear decay process.
A radiative nucleus may emit an electron from $i$-th atomic shell with branch ratio $\kappa_i$, 
or a $\gamma$ photon with branch ratio $\kappa_\gamma$.
Internal conversion coefficient $\alpha_T$ is defined as\cite{ICC-BAND2002}
\begin{equation}
\alpha_T=\sum_i\alpha_i=\sum_i\frac{\kappa_i}{\kappa_\gamma}
\end{equation}
For many nuclear isomers, $\alpha_T$ may much larger than 1, $\alpha_T\gg 1$, 
which means the nucleus is very difficult to decay through the channel $N^*\rightarrow N+\gamma$,
and prefers the channel $N^*\rightarrow N+e$.

Therefore, as the inverse processes, 
for $\alpha_T\gg 1$ nuclei, $e+N\rightarrow N^*$ is much higher efficient way to excite  the nuclei $N$ than  
$\gamma+N\rightarrow N^*$.
If the original electron was free, and then was bounded to the atom  (Fig.\ref{fig.NEEC}),
\begin{equation}\label{eq.neec}
N+e^{(free)}\rightarrow N^*+e^{(bounded)},
\end{equation}
the process is called nuclear excitation by electron capture (NEEC)\cite{NEEC-theory-PhysRevA.73.012715,NEEC-theory-PhysRevA.75.012709}.
If the original electron was bounded, and then was bounded to a deeper level in the atom (Fig.\ref{fig.NEET}),
\begin{equation}\label{eq.neet}
N+e^{(bounded1)}\rightarrow N^*+e^{(bounded2)},
\end{equation}
the process is called nuclear excitation by electron transfer (NEET)\cite{NEET-PRL2000-Au197, NEET-Os189-PRC2000, NEEC-NEET-MOREL2004608,NEET-Morita1976}.

The NEEC process was first proposed by Goldanskii \cite{NEEC-Goldanskii1976}
Based upon the Femi Golden rule, one may writte the NEEC cross section as
\cite{NEEC.NEET.AIP.Conf.1.1945195, NEEC-theory-PhysRevA.73.012715,NEEC-theory-PhysRevA.75.012709},
\begin{equation}\label{eq.NEEC}
\sigma_{NEEC}(E)=\frac{(2\pi\hbar)^2}{2m_e E}\frac{2J_f+1}{2J_i+1}\Gamma^e_{J_i\rightarrow J_f}
\frac{\Gamma^{tot}}{(E-E_r)^2+\left(\frac{\Gamma^{tot}}{2}\right)^2},
\end{equation}
where $m_e$ is the mass of the electron,
$J_i$ and $J_f$ are initial and final nuclear spin respectively,
$\Gamma^{tot}$ is the total transition width,
$\Gamma^e_{J_i\rightarrow J_f}$ is the transition width of $J_i\rightarrow J_f$,
$E$ is the energy,
and $E_r$ is the resonance energy.

The NEET process was first proposed by Morita in 1976\cite{NEET-Morita1976}.
In weak coupling limit ($\kappa\rightarrow 0$), the NEET cross section can be written as
\cite{NEET-Os189-PRC2000},
\begin{equation}\label{eq.NEET}
\sigma_{NEET}^{\kappa\rightarrow 0}=\frac{\Gamma_i\Gamma_f}{\Gamma_i}
\frac{\kappa^2}{(E_i-E_f)^2+\left(\frac{\Gamma^{tot}}{2}\right)^2},
\end{equation}
where $\kappa=\braket{f|i}$,
the $i$ and $f$ represent the initial and final states respectively,
$\Gamma$ is the transition width,
$E$ is the bounding energy.

In a laser-plasma environment where a lot of energetic electrons are around, 
NEEC and NEET may dominate the processes of generating nuclear isomers 
through reactions shown in Eq.\ref{eq.neec} and \ref{eq.neet}.


\begin{figure}
 \centering

 \begin{tikzpicture}[scale=0.2]
 \pgfmathsetmacro{\L}{8}
 \pgfmathsetmacro{\H}{6}
 \pgfmathsetmacro{\Gap}{1.5}
 \draw[thick, black] (0,0) -- (\L,0);
 \draw[thick, blue] (\L+\Gap,0) -- (2*\L+\Gap,0);
 \draw[thick, black] (2*\L+5*\Gap,0) -- (3*\L+5*\Gap,0);
 \draw[thick, blue,dotted] (3*\L+6*\Gap,0) -- (4*\L+6*\Gap,0);
 \draw[thick, fill=white,font=\fontsize{6}{6}\selectfont] (2.5*\L+5*\Gap,0) circle (0.5) node {e};
 \draw[thick, fill=red,font=\fontsize{6}{6}\selectfont] (1.5*\L+1*\Gap,0) circle (0.6) node {N};

 \draw[very thick, blue,dotted] (2*\L+3*\Gap,-0.5*\H) -- (2*\L+3*\Gap,1.8*\H);
	
 \draw[thick, black] (0,\H) -- (\L,\H) ; 
 \draw[thick, blue,dotted] (\L+\Gap,\H) -- (2*\L+\Gap,\H);
 \draw[thick, black] (2*\L+5*\Gap,\H) -- (3*\L+5*\Gap,\H);
 \draw[thick, blue] (3*\L+6*\Gap,\H) -- (4*\L+6*\Gap,\H);
 \draw[thick, fill=white,font=\fontsize{6}{6}\selectfont] (\L/2,\H) circle (0.5) node {e};
 \draw[thick, fill=red,font=\fontsize{6}{6}\selectfont] (3.5*\L+6*\Gap,\H) circle (0.6) node {N};
 \node at (0.5*\L,1.5*\H) {Atomic};
 \node at (2.5*\L+5*\Gap,1.5*\H) {Atomic};
 \node at (1.5*\L+\Gap,1.5*\H) {Nuclear};
 \node at (3.5*\L+6*\Gap,1.5*\H) {Nuclear};
 \draw[snakeline] (\L/2, \H-0.6) -- (0.5*\L,0);
 \draw[snakeline] (1.5*\L+\Gap, 0.6) -- (1.5*\L+\Gap,\H);

 \node at (\L+0.5*\Gap,-0.6*\H) {Initial State};
 \node at (3*\L+6.5*\Gap,-0.6*\H) {Final State};

 \end{tikzpicture}	

\caption{
Atomic and nuclear states involved in a NEET transition.
In the initial state, the electron is on a bounded excited state and the nucleus on the ground state;
while in the final state, the electron is on a bounded lower energy state and the nucleus on an excited state.
}

 \label{fig.NEET}
\end{figure}
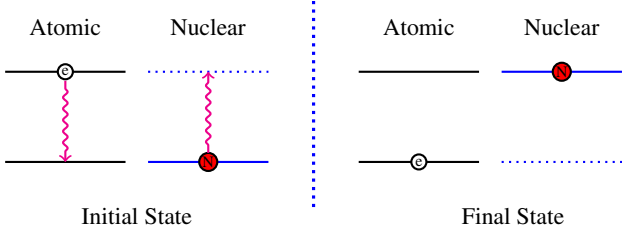

\subsection{Electron Bridge}

The same as in the NEET and NEEC processes, the electronic bridge process can be the dominant channel for a nuclear isomer which has a resonance channel available \cite{EB-Krutov-1968, EB-PRL.124.192502}

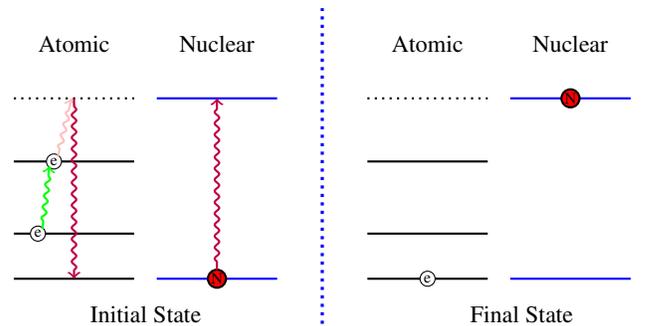
\begin{figure}[b]
 \centering
\begin{tikzpicture}[scale=0.2]
 \pgfmathsetmacro{\L}{8}
 \pgfmathsetmacro{\H}{6}
 \pgfmathsetmacro{\Gap}{1.5}
 \draw[thick, black] (0,0) -- (\L,0);
 \draw[thick, blue] (\L+\Gap,0) -- (2*\L+\Gap,0);
 \draw[thick, black] (2*\L+5*\Gap,0) -- (3*\L+5*\Gap,0);
 \draw[thick, blue] (3*\L+6*\Gap,0) -- (4*\L+6*\Gap,0);
 \draw[fill=white,font=\fontsize{6}{6}\selectfont] (2.5*\L+5*\Gap,0) circle (0.5) node {e};
 \draw[thick, fill=red,font=\fontsize{6}{6}\selectfont] (1.5*\L+1*\Gap,0) circle (0.6) node {N};

 \draw[very thick, blue,dotted] (2*\L+3*\Gap,-0.5*\H) -- (2*\L+3*\Gap,3*\H);

 \draw[thick, black,dotted] (0,2*\H) -- (\L,2*\H) ; 
 \draw[thick, blue] (\L+\Gap,2*\H) -- (2*\L+\Gap,2*\H);
 \draw[thick, black] (0,0.5*\H) -- (1*\L+0*\Gap,0.5*\H);
 \draw[thick, black] (0,1.3*\H) -- (1*\L+0*\Gap,1.3*\H);
 \draw[thick, black] (2*\L+5*\Gap,1.3*\H) -- (3*\L+5*\Gap,1.3*\H);
 \draw[thick, black] (2*\L+5*\Gap,0.5*\H) -- (3*\L+5*\Gap,0.5*\H);
 \draw[thick, black,dotted] (2*\L+5*\Gap,2*\H) -- (3*\L+5*\Gap,2*\H);
 \draw[thick, blue] (3*\L+6*\Gap,2*\H) -- (4*\L+6*\Gap,2*\H);
 \draw[fill=white, font=\fontsize{6}{6}\selectfont] (\L/3,1.3*\H) circle (0.5) node {e};
 \draw[fill=white, font=\fontsize{6}{6}\selectfont] (\L/5,0.5*\H) circle (0.5) node {e};
 \draw[snakeline,pink] (\L/3+0.25,1.3*\H+0.3) -- (0.48*\L,2*\H);
 \draw[snakeline,green] (\L/5+0.3, 0.5*\H+0.3) -- (\L/3-0.3,1.3*\H-0.3);
 \draw[snakeline, purple] (0.5*\L,2*\H) -- (0.5*\L,0);
 \draw[thick, fill=red,font=\fontsize{6}{6}\selectfont] (3.5*\L+6*\Gap,2*\H) circle (0.6) node {N};
 \draw[snakeline, purple] (1.5*\L+1*\Gap, 0.6)  -- (1.5*\L+1*\Gap,2*\H) ;

 \node at (0.5*\L,2.6*\H) {Atomic};
 \node at (2.5*\L+5*\Gap,2.6*\H) {Atomic};
 \node at (1.5*\L+\Gap,2.6*\H) {Nuclear};
 \node at (3.5*\L+6*\Gap,2.6*\H) {Nuclear};

 \node at (\L+0.5*\Gap,-0.4*\H) {Initial State};
 \node at (3*\L+6.5*\Gap,-0.4*\H) {Final State};

 \end{tikzpicture}	

\caption{
Atomic and nuclear states involved in an EB transition.
In the initial state, the electron is at a bounded state and the nucleus on its ground state.
When the electron absorbs a photon (pink line) or even photons (green+pink lines), 
it is excited to a virtual higher state.
The electron may drop to the ground state, while at the same time the nucleus is excited to a higher energy state.
}

 \label{fig.EB}
\end{figure}

The EB process is shown in Fig.\ref{fig.EB}.
In the beginning, the electron involved is at a lower bounded state with energy $E_{e0}$, and the nucleus is in the ground state.
The electron may absorb one or even two photons with energies $E_{p1}$ (and $E_{p2}$ if two photons are absorbed) and jumps to a virtual state $E_e^{*}$.
If $E_{e0}+E_{p1}+E_{p2}=E_e^{*}=E_m$, where $E_m$ is the excited energy of the nuclear isomer,
the nucleus can jump up to the corresponding isomer state resonantly.

The EB has the following advantages: 
First, isomers which relatively larger $E_m$ can be studied with help of the EB process. 
Currently, optical lasers with photon energies of $E_p$ larger than tens eV are not available. 
Free Electron Lasers (FELs) may have higher photon energies but have much larger bandwidths\cite{FEL-McNeil-2010NatP}. 
By choosing a proper $\Delta E=E_{p1}+E_{p2}$, 
one can study a higher $E_m$.
Secondly, the EB can highly improve the photon absorption cross section.
In fact, even if one has a laser with a proper laser beam which has photon energy $E_p=E_m$, 
the photon absorption cross section  $\sigma_\gamma$ of the reaction $\gamma+X\rightarrow X^{(*)}$ is still very small,
mainly because of the very small nuclear diameter. 
However, with the EB machinisium, the enhancement factor 
 can be increased by a factor between 10 to $10^9$\cite{Th229-Wense2020EPJA, CaiNengQiang2021}, 
The enhance factor for single electron bridge can be written as\cite{Th229-Wense2020EPJA}, 
\begin{equation}\label{eq.EB}
R_{eb}\equiv\frac{\sigma_{eb}}{\sigma_\gamma}\propto 
\frac{\Gamma_{eb}\Gamma_l^2}{\Gamma_\gamma}
\end{equation}
where $\sigma_{eb}$ is the cross section through the EB mechanism, 
$\Gamma_l$ is the linewidth of the laser,
$\Gamma_{eb}$ is the nuclear EB channel decay width,
and $\Gamma_\gamma$ is the nuclear $\gamma$-decay channel  width.
For the double photon EB case, one has the similar dependence as that in one-photon EB cases\cite{CaiNengQiang2021}.

\section{Potential Applications related to FNMS}
\subsection{Nuclear Clock}
Clocks are fundamental devices for physics.
No precise clocks, no modern physics. 
From ancient sundials and sandglass to modern atomic clocks, 
more and more accurate clocks have been made in history,
and the expectation of an even more precise one is never stopping.
In 2019,  scientists from  the National Institute of Standards and Technology demonstrated an Al$^+$ clock with a total uncertainty of 9.4$\times$10$^{-19}$ s\cite{Clock-Al-PRL2019}. 

Normally, a nuclear transition has a much smaller $\Delta E/E$, where the $\Delta E$ is spectral linewidth and $E$ is the energy difference of the transition.
A smaller $\Delta E/E$ means a higher accurate clock\cite{Th229.Clock.Nat.Rev.2021}. 
Recently, a transition in cesium atoms was founded to have the uncertainty of  $2.5\times 10^{-19}$ s\cite{Clock-Sr87-PRL2018Marti}.
However, it is more and more difficult  to find a transition with an uncertainty smaller than the order of $10^{-19}$ s in atoms.
While in nuclear isotopes, there are many transitions in which $\Delta E/E$  is much smaller than these in atoms.
it is expected that a new generation of clocks will base on nuclear transitions, i.e. nuclear clock (or gamma clock).

There have two nuclear isomers, $^{229\text{m}}$Th and $^{235\text{m}}$U, which are thought to be promising for producing a nuclear clock. 
The isomer $^{229m}$Th has an excited energy of  $(8.10\pm0.17) $ eV
\cite{Th229-lifetime-PRL2020, Th229-Peik-2003, Th229-PRL2012-Campbell,Th229-Nat.Wense-2016}.  
The isomer $^{235m}$U has a higher excited energy of 76 eV
\cite{U235-EB-PRL2018Berengut}.
Their excited energies are relatively low and are reachable by the current optical laser technologies.


How to excited nuclei to their isomer states is the precondition of making a nuclear clock. 
Limited by the laser technologies, lasers with proper wavelengths and precision to match the isomeric energies are not available today. 
Therefore,  indirect excitation schemes such as NEEC, NEET, and EB  are investigated as alternate ways.

\subsection{Nuclear Battery}

The FNMS physics has another application, nuclear batteries.
Due to their long-lasting time,  environmental stability,
 and high energy densities,
nuclear batteries have been widely used in aerospace, deep ocean, polar areas, cardiac pacemakers, and micro-electromotor, etc\cite{Nucl.Battery.Rev.2014Prelas,Nucl.Batt.2019CPL}.
Many radiative materials can be used to make nuclear batteries, the ones with $\alpha$, $\beta$, or $\gamma$ decays.
Nuclear isomers are among them and have advantages  like  the potential of rechargeability.
$^{93m}$Mo,  $^{180m}$Ta, and $^{178m}$Hf are candidates for nuclear battery production.
The processes of NEET, NEEC, and EB may be used for enhancing the production rates.

\section{High intensity lasers and FNMS}

High intensity laser facilities provide new opportunities to study various physical processes happening on FNMS, especially nonlinear processes.

Most of the nucleus-laser interactions with today's laser techniques are indirect.
The record laser intensity is  $1.1\times 10^{23}$ W/cm$^2$\cite{LaserE23-2021}.
The laser's electron field $\mathbf{E}$ can be written as 
\begin{equation}
\mathbf{E}=27.4\times \left(\frac{I}{[W/cm^2]}\right)^{1/2} [V/cm].
\end{equation}
As a classic limit estimation, 
the order of the energy $\mathcal{E}$ acquired by a proton inside a nucleus  due to the electric field can be estimated by,
\begin{equation}\label{eq.E.e.D}
\mathcal{E}=\mathcal{O}(De\mathbf{E}),
\end{equation}
where $D$ is the nuclear diameter, and $e$ is the charge of a electron.
If take $D\approx 10$ fm,  $\mathcal{E}=10$ eV,
One can see that low lying levels like those in $^{229}$Th(8.1 eV) and $^{235}$U(74 eV)
may have the possibility to be excited directly by the strongest laser today.
However, most nuclear excited states are in keV and MeV ranges. 
Until the Schwinger limit\cite{Schwinger-Limit-PRL2003Bulanov}, corresponding to $2.3\times 10^{29}$ W/cm$^2$, or $E=1.3\times 10^{16}$ V/cm, being achieved,
most nuclear isotopes cannot be excited directly by lasers.

Indirect nucleus-laser interaction could be huge due to the resonance mechanisms of NEEC, NEET, and EB.
For the NEEC and NEET processes, electrons with energy of keV order are needed.
One can estimate the order of electron energy by using a similar formula as Eq.\ref{eq.E.e.D},
\begin{equation}\label{eq.E.e.lambda}
\mathcal{E}_e=\mathcal{O}(\lambda e\mathbf{E}),
\end{equation}
where $\lambda$ is the laser wavelength.
The physical meaning of $\mathcal{E}_e$ is the  energy order of one electron driven by a laser during one circle of the laser's electromagnetic field oscillation.
In fact, in a typical high intensity laser experiment, electron temperature can easily reach keV level\cite{DD-zhang2017deuteron}.
At the same time, atoms are ionized to high charge states. 
Therefore, a laser induced plasma  is an ideal platform for the NEEC and NEET studies\cite{NEEC-FEL-Gunst2015,HIL-HeYF2020}.

Furthermore, as shown in Eq.\ref{eq.NEEC}, \ref{eq.NEET}, and \ref{eq.EB},
the NEEC, NEET, and EB are highly dependent on the decay width of the nuclei.
In plasma or high-intensity laser fields,
without the shielding from the electron cloud,
the nuclei can experience a relatively strong time-dependent potential $\Delta V(t)$.
This extra $\Delta V(t)$ may cause broader bounded state width.
Considering the fact that many nuclear isomers' energy widths are much smaller than $eV$,
an eV level extra potential $\Delta V(t)$, as estimated by Eq.\ref{eq.E.e.D}, 
could improve absorbing widths dramatically.

Normally, for the EB processes, narrow band photon sources are needed.
However, in laser induced plasma, the atomic, as well as nuclear, absorption lines are broadened due to the Doppler effect\cite{Charge-Dopl-PRL2004Foord,Doppler-Plasma-JOSA1967,Doppler-e-ion-JPCM2000Chihara, LaserBroading2019CPC}.
Furthermore, one nonlinear process, surface enhancement effect\cite{Surface-enhance-RevModPhys.57.783},  
may also be used to improve the EB effects when using high intensity lasers.
It is well known that, with the surface-enhanced Raman scattering, Raman spectra have improved sensitivity over a factor of $10^{10}$ than ordinary ones, and a single molecule can be detected by using the method\cite{Surf-Raman-PRL1997Kneipp}.
This nonlinear effect can also improve the NEEC, NEET, and EB processes in laser-induced plasmas.

\section{summary}
In summary, we review the puzzles which may relate to FNMS.
The puzzles include so-called the Deep Dirac Level puzzle, the neutron lifetime puzzle, and the proton size puzzle.
It was suspected that some unknown physical processes on FNMS may exist, and to be discovered.
We also review the mechanisms linking nuclear decays to atomic structures, the Nuclear Excitation by Electron Capture, the Nuclear Excitation by Electron Transfer, and the Electron Bridges.
For a radiative nucleus having $\beta$-related decay channels, 
the weak force connects the nucleus and the electrons together, 
and the NEEC, NEET, and EB are the right ways to study them.
From the experimental point of view, 
the high intensity lasers today are still not strong enough to change nucleons in a nucleus directly.
However, all electrons bounded to a nucleus can be affected by high intensity lasers today.
Considering the role of electrons in weak interaction, 
the high intensity laser facilities may play be critical in studies of the puzzles on FNMS.

\section{Acknowledgement}
This work is supported by 
the National Nature Science Foundation of China (under grant no. 11875191),
and the Strategic Priority Research Program of the Chinese Academy of Sciences (grant no. XDB16).


\end{CJK*}
\end{document}